\documentclass[fleqn,usenatbib]{mnras}

\usepackage{graphicx}   % Figuras
\usepackage{amsmath}    % Ecuaciones
\usepackage{amssymb}    % Símbolos matemáticos
\usepackage{newtxtext,newtxmath} % Tipografía recomendada
\usepackage{booktabs}
\usepackage[normalem]{ulem}
\usepackage{float}
\usepackage{caption}
\title[Long-Term Analysis of PKS\,2155$-$304]{PKS\,2155$-$304: Long-Term Optical Photometric Monitoring and Variability Analysis}

\author[Weiss et al.]
{
J. L. Weiss$^{1,2}$\thanks{E-mail: jweiss@iar.unlp.edu.ar},
I. Andruchow$^{1,2}$,
E. J. Marchesini$^{3}$,
S. Cellone$^{1,4}$,
L. Zibecchi$^{1,5}$,
J. P. Varela$^{1}$,
\newauthor
L. Mammana$^{1,4}$,
E. Jofré$^{6,7}$ and
R. Petrucci$^{6,7}$
\\
$^{1}$ Facultad de Ciencias Astron\'omicas y Geof{\'\i}sicas  (FCAG), Universidad Nacional de La Plata, Paseo del Bosque s/n, FWA, B1900, La Plata, Provincia\\ de Buenos Aires, Argentina\\ $^{2}$ Instituto Argentino de Radioastronom\'ia, CONICET--CICPBA--UNLP, CC5 (1894), Villa Elisa, Provincia de Buenos Aires, Argentina\\
$^{3}$ Osservatorio di Astrofisica e Scienza dello Spazio di Bologna, Istituto Nazionale Di Astrofisica (OASBo-INAF), Via Gobetti 93/3, I-40129, Bologna, Italy\\
$^{4}$ Complejo Astronómico El Leoncito, CONICET--UNLP--UNC--UNSJ, Av. Argentina 147 norte, J5400FJB, San Juan, Argentina\\
$^{5}$ Instituto de Astrofísica de La Plata (IALP), CONICET-UNLP, B1900FWA La Plata, Provincia de Buenos Aires, Argentina\\
$^{6}$ Universidad Nacional de Córdoba - Observatorio Astronómico de Córdoba, Laprida 854, X5000BGR, Córdoba, Argentina\\
$^{7}$ Consejo Nacional de Investigaciones Científicas y Técnicas (CONICET), Godoy Cruz 2290, CPC 1425FQB, CABA, Argentina}

\date{Accepted XXX. Received YYY; in original form ZZZ}

\pubyear{2025}

\begin{document}

\maketitle

\begin{abstract}
Through the detailed study of the optical flux behaviour in blazars over time, it is possible to infer the conditions responsible for their observed emission. PKS\,2155$-$304, a BL\,Lac object detected from radio to TeV energies, is among the brightest blazars in the southern hemisphere. We present optical monitoring spanning over two decades using telescopes at Complejo Astronómico El Leoncito and Estación Astrofísica de Bosque Alegre, Argentina. Differential light curves in the $B$, $V$, $R$, and $I$ bands reveal significant variability on weekly and longer timescales, with occasional changes on sub-four-hour scales. The optical spectral index remained negative, consistent with non-thermal emission, and hardened over the past nine years. Evidence for quasiperiodic behaviour on 20–30 day timescales was found, while correlations with X-ray fluxes were weak, suggesting distinct emission components in the two bands. These results highlight the pronounced optical variability of PKS\,2155$-$304 and provide insight into its multi-band emission mechanisms.
\end{abstract}


%%%%%%%%%%%%%%%%%%%%%%%%%%%%%%%%%%%%%%%%%%%%%%%%%%
\section{Introduction}
    \label{sec:Introduction}
   Active galaxies are objects that exhibit significant energy emission in their nuclear regions, with bolometric luminosities on the order of $10^{42}$\,erg\,s$^{-1}$, potentially reaching values up to $10^{48}$\,erg\,s$^{-1}$ \citep{fan}. These values cannot be explained by the known emission mechanisms in most galaxies (stars, HII regions, supernova remnants, etc.). The emission from the active nuclei of galaxies (AGN) is distributed across the entire electromagnetic spectrum \citep{peterson}. The power emitted by them usually varies on time scales of years, reaching scales of days, hours, and even minutes \citep{romero}, depending on their class. This rapid variability points to a highly energetic process more efficient than nuclear fusion. AGN are among the most luminous extragalactic sources and represent a significant fraction of the universe's electromagnetic energy production \citep{krawczynski}.

   According to the unified model \citep{lynden, urrypadovani}, blazars are those AGNs where the line of sight is close to the direction of the jet. Their emission is dominated by relativistic jets of high-energy particles. Their Spectral Energy Distribution (SED) is characterized by two non-thermal components, resulting in a distribution with two peaks: the first occurs in the radio/UV region and is associated with synchrotron emission, while the second lies between $X$-Rays and $\gamma$-Rays. In the leptonic model, this high-energy component is attributed to the inverse Compton effect \citep[e.g.,][]{ghisellini1985, marscher, bloom, chiadberge, paiano}, where low-energy photons are upscattered by the same ultrarelativistic electrons in the jet. Alternatively, in the hadronic model, the high-energy emission is dominated by the synchrotron radiation of ultrarelativistic protons \citep[e.g.,][]{dermer, sikora, blazejowski, ghisellini2009}. A key characteristic of blazar emission is its pronounced temporal variability across multiple wavelengths, offering valuable insights into the microphysical processes occurring within the jet. Notably, blazars occasionally exhibit significant flux increases, known as flares, which have been observed in various sources over time \citep[e.g.][]{abramowski, hong, escudero}. In particular, observing these objects in the $V$ and $R$ bands allows the study of the variability of their emission in the optical range, which may be related to variability at other wavelengths, including $\gamma$-rays and $X$-rays. Furthermore, using optical observations, it is possible to characterize the synchrotron emission of the object \citep{maraschi, dermer}.
   
   The source known as PKS\,2155$-$304, located at $z = 0.116$ \citep{falomo}, is a BL\,Lac$-$type blazar. This object is frequently observed, as it is one of the brightest blazars in the southern sky and exhibits strong variability across all wavelengths. In the optical band, it has shown short-term variability on timescales as brief as 15 minutes \cite{paltani}, and quasi-periodicities of approximately 0.7 days have been reported \citep{urry}. At high energies, it has been monitored since the early TeV detection experiments \citep{chadwick}, being the first source to exhibit variability on timescales of hours in this energy range \citep{aharonian}, thus making it a key target for high-energy observations.

    Multiple studies in the literature have searched for possible quasi-periodicities in the light curves of this object, yielding a variety of results. Long-term quasi-periodic oscillations (QPOs) have been reported, such as those of $\sim$315 and $\sim$620 days by \cite{Sandrinelli18}, $\sim$318 days by \cite{Sandrinelli16}, and $\sim$317 days by \cite{zhang}. \cite{bhatta20} also reported a QPO at approximately $\sim$649 days. Similar long-term signals (e.g., $\sim$300 and $\sim$700 days) have been discussed in works such as \cite{chevalier15}. On the other hand, \cite{paltani} found that optical variability can occur on timescales ranging from 10 to 40 days. In contrast, other studies have found no evidence of persistent periodicities; for instance, \cite{chatterjee} reported no characteristic timescales in their multiwavelength analysis, while \cite{covino} concluded that no periodic behavior is present in the optical band.
    
    In this work, we characterize the optical variability of this source by studying its light curves. For this purpose, we use the data archive of the CONGA group (\textit{Caracterización Observacional de Núcleos de Galaxias Activas}, FCAG, UNLP – CO\-NI\-CET), which includes observations collected at Argentine astronomical facilities over the past two decades. In addition, we incorporate new data obtained by our team over the last two years. In order to have a better understanding of the physical phenomena occurring in the source, we analyze the variability state on different time scales, and we search for possible quasi-periodicities. Using available data in public databases, we study the presence of correlations between the fluxes in the $X$-rays and optical bands. 

    This paper is organized as follows: in Section 2 we describe the observational data, in Section 3 we present the statistical tests used, and in Section 4 we show the results of the behaviour of the light curves. The spectral index behaviour and the search for quasiperiodicities are discussed in Section 5, and finally, we present the summary and conclusions in Section 6.

\section{Observations and data reduction}

The CONGA group has access to optical images of PKS\,2155$-$304 taken at Argentinian astronomical facilities since July 1997. For this study we used photometric data in the $B$, $V$, $R$, and $I$ bands (Johnson-Cousins filter system; \citealp{bessel}). Observations were conducted primarily with the 2.1\,m Jorge Sahade Telescope (JS), with additional data collected using the 0.6\,m Helen Sawyer Hogg telescope (HSH) ---both at the Complejo Astronómico El Leoncito (CASLEO), in San Juan---, and the 1.5\,m telescope at the Estación Astrofísica de Bosque Alegre (EABA) in C\'ordoba. 
For a single night in 1997, the detector was a Tek$-$1024 CCD, with a gain of 1.98 e${^-}$/ADU and a readout noise of 9.6 e${^-}$. In subsequent years, the CCD Roper Versarray-2048 was used, with a gain of 2.18\,e${^-}$/ADU and a readout noise of 3.11\,e${^-}$. The exposure times varied depending on the filters used, the state of the source, and the sky conditions. A summary of the minimum, maximum, and average exposure times for each photometric band, grouped by telescope, can be found in Table \ref{tab:tiempos}.

\begin{table}
\centering
\caption{Telescopes and photometric bands are listed in the first and second columns, respectively. The third, fourth, and last columns indicate the minimum, mean, and maximum exposure times (in seconds), in that order.}
\begin{tabular}{c c c c c}
\toprule
Telescope & Band & $t_\text{min}$ & $t_\text{average}$ & $t_\text{max}$  \\
 & & [s] & [s] & [s] \\
\midrule
JS & $B$ & 60 & 192 & 600  \\
& $V$ & 20 & 113 & 720  \\ 
& $R$ & 10 & \phantom{1}74 & 600  \\ 
& $I$ & 16 & 117 & 500  \\ 
\midrule
HSH & $V$ & 20 & 479 & 690  \\ 
& $R$ & 240 & 443 & 630  \\ 
& $I$ & 300 & 403 & 570  \\ 
\midrule
EABA & $B$ & 40 & 89 & 200  \\
& $V$ & 20 & 37 & 150  \\ 
& $R$ & 10 & 24 & 80  \\ 
& $I$ & 25 & 37 & 100  \\ 
 \bottomrule
\end{tabular}
\label{tab:tiempos}
\end{table}

Our research group initially obtained observational data spanning from 1997 to 2022, consisting of a single night of observations in 1997 and a continuous set of observations from 2015 onward, with only the data from 1997 and 2015 having been previously published \citep{romero, 2002A&A...390..431R, zibecchi2017, zibecchi2024}. To expand the dataset and achieve broader temporal coverage, additional observations were conducted using the JS telescope in 2023 and 2024. As a result, the final dataset comprises over 6,500 images collected between 1997 and 2024, spanning 95 observation nights. All optical images were reduced using {\sc IRAF}\footnote{\textsc{iraf} is distributed by the National Optical Astronomy Observatories, which are operated by the Association of Universities for Research in Astronomy, Inc., under cooperative agreement with the National Science Foundation.}, by means of tasks developed by our research group in order to optimize the reduction process.

We obtained $X$-ray data from the Swift-XRT\footnote{https://www.swift.ac.uk/LSXPS/} catalog \citep{evans}, considering only observations from 2015 to 2023, as this period corresponds to the majority of the group's optical data. The data were subdivided into their hard (2–10 keV), medium (1–2 keV), and soft (0.3–1 keV) components. In addition, we considered the totality of data points in the entire $X$-ray range, i.e., from 0.1 keV to 10 keV, a band to which we will henceforth refer to as the total $X$-ray band.

\subsection{Differential photometry}

Differential photometry consists on comparing the flux of an object of interest with that of one or more nearby reference stars in the field of view. In this way, relative brightness changes can be detected, instead of measuring the absolute flux of the object of interest. %
%The main \sac{advantage} of differential photometry \sac{when studying} brightness variations in a given object \sac{is that systematic effects arising on atmospheric and instrumental issues are (mostly) suppressed}.
For this, it is necessary that the reference stars be non-variable. This technique has been widely used for AGN variability analysis \citep[e.g.,][]{andruchow, gopal} and, in particular, has been applied to the case of PKS\,2155$-$304 \citep[e.g.,][]{mantegazza, zibecchi2017, zibecchi2024}.

In our implementation, differential photometry involves measuring the instrumental magnitudes of three objects: the object of study, which we will call \textbf{V}, a comparison object \textbf{C}, and a control object \textbf{K}. \textbf{C} and \textbf{K} should be objects that are close to \textbf{V} in the field so that the photometric information of these three objects can be obtained with a single exposure.

This technique, in principle, allows for the elimination of atmospheric effects and variable observing conditions, as it is assumed that the reference stars are observed under the same conditions as the object of interest. Any variation due to sky fluctuations will thus similarly affect the three sources.

Once the three sources are measured, we proceed to obtain the difference in instrumental magnitudes  $m_{v,\mathrm{ins}}-m_{c,\mathrm{ins}}$ and $m_{k,\mathrm{ins}}-m_{c,\mathrm{ins}}$. The latter is important because it allows us to detect any possible variability in the comparison object and/or the control object. It is also a measure of the intrinsic instrumental precision and provides a comparison to determine if the light curve of the source is variable or not (e.g., \citealp{cellone}).

A key strength of this method is that it delivers consistent results even in the absence of strictly photometric conditions. According to the criterion of \cite{howell1986}, it is recommended that the magnitude of the comparison star be similar to that of the source, while the magnitude of the control star can be somewhat brighter than the other two.

Furthermore, knowing the magnitude of \textbf{C}  ($m_{c,\mathrm{std}}$) in the standard system, it is possible to obtain the magnitude of the object \textbf{V} in the standard system, using the following equation:
\begin{equation}
    m_{v,\mathrm{std}}=(m_{v,\mathrm{ins}}-m_{c,\mathrm{ins}})+m_{c,\mathrm{std}}.
    \label{sistema estandar}
\end{equation}

\begin{figure}
    \centering
    \includegraphics[width=.47\textwidth]{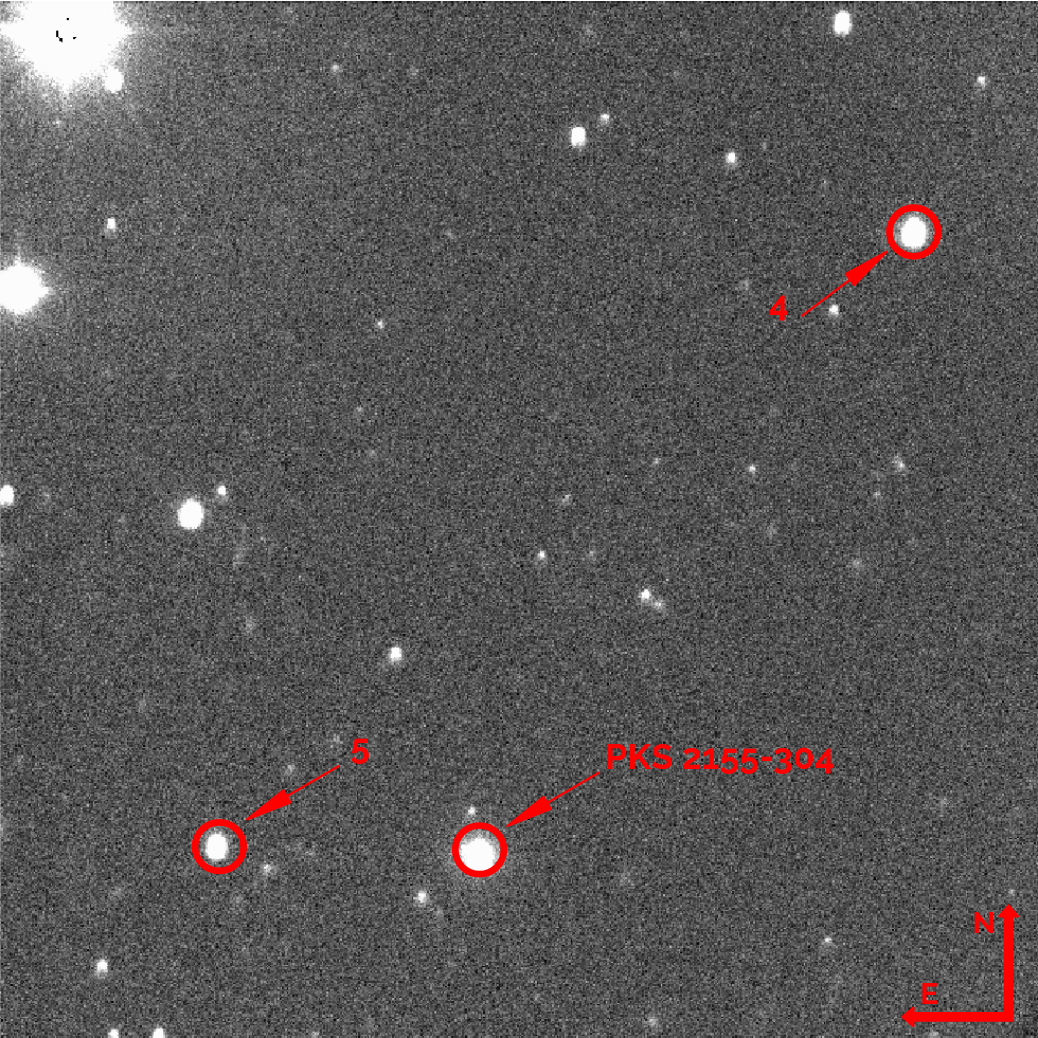}
    \caption{Field of PKS\,2155$-$304 with comparison objects 4 and 5 from \protect\cite{hamuy}. The field of view is $5 \times 5$ arcmin.}
    \label{fig:campo}
\end{figure}

In this study, we used multiple sets of comparison and control objects, as the blazar exhibited significant flux variations during the observed period. This set of stars was chosen to ensure a stable and reliable photometric baseline. Additionally, this set was used to obtain the magnitude of the blazar in the standard system. Specifically, our analysis was based primarily on comparison object 5 from \cite{hamuy}, with object 4 from the same publication being employed in some cases. Figure \ref{fig:campo} shows the field of PKS\,2155$-$304 with comparison objects 4 and 5 from that publication.

\section{Statistical tests}\label{sec:std}

 To determine whether PKS\,2155$-$304 exhibited significant variations across different timescales, we applied two statistical tests to the obtained differential light curves: Fisher’s $F$-test \citep{de_Diego} and the $C$ criterion \citep{howell1988}. These tests are widely used to assess variability in AGNs (e.g. \citealp{gupta}, \citealp{singh}, \citealp{ege}) Both tests rely fundamentally on the standard deviation of the 'object-comparison' light curve ($\sigma_1$) and that of the 'comparison-control' light curve ($\sigma_2$), the latter serving as a measure of the quality of the night. These tests were chosen due to their robustness in distinguishing intrinsic variations from measurement uncertainties. The $C$ criterion provides a straightforward statistical comparison between the variability of the target source and the expected photometric and instrumental noise, allowing us to identify significant deviations \citep{romero}. In addition to the $C$ criterion, we also applied the $F$-test because statistical analyses of variability often yield contradictory or diverse results \citep{zibecchi2017}. Ideally, the classification of a source's variability state should not depend on the specific statistical method employed. By using both tests, we aim to ensure a more consistent and reliable assessment of the blazar’s variability.

To account for the difference in magnitudes between the science object and the control and comparison objects, we followed the procedure as detailed in \cite{zibecchi2017} implementing the weight factor $\Gamma$ \citep{howell1988}. We will refer to the mathematical value of the weighted parameter of the $F$-test as $F_\Gamma$ and that of the $C$ criterion as $C_\Gamma$. 

Due to the greater number of photometric points in the $V$ and $R$ bands than in the $B$ and $I$ bands, we performed the short-term statistics using only the data in $V$ and $R$. On the other hand, we used the measurements in $B$ and $I$, mainly for statistical analysis on a monthly and yearly scale. 

We used a 99.5\% confidence level to determine whether a light curve, in a given band and timescale, is variable.

\subsection{Z-transformed Discrete Correlation Function}

To analyze possible correlations, we used the Z-transformed Discrete Correlation Function (ZDCF), which is especially useful for the robust estimation of the cross-correlation function of sparsely and unevenly sampled time series. In particular, it is widely used to correlate different bands in AGNs \citep[see for example:][]{shapalova, jankov, kovacevic}. This algorithm differs from the Discrete Correlation Function (DCF) in that it groups data points into equally populated bins and uses Fisher's Z-transformation to stabilize the highly skewed distribution of the correlation coefficient. Simulations show that, according to various criteria, these modifications improve the performance of ZDCF over that of DCF \citep{tal}.

\section{Results}

\subsection{Variability}

In this section, we present the results of the variability behaviour of the source, at different timescales (annual, monthly, intraday, and microvariability). Tables~\ref{tab:teststotal}, \ref{tab:testsmensual}, and \ref{tab:testsinterdia} follow the same format: the first column lists the analysed time period, the second the photometric band, the third the number of observation nights, the fourth the total number of photometric points in the interval considered, $C_\Gamma$ is the value of the weighted $C$ parameter, $F_\Gamma$ the value of the weighted $F$ parameter, and the last column shows the dispersion of the differential control light curve, $\sigma_2$, which we adopted as a measurement of the observational error. As described in Section \ref{sec:std}, we employ the $\Gamma$ factor to account for magnitude differences between the comparison stars and the target, ensuring the dispersion reflects the observational error for the source. Results of positive variability are shown in bold.

\subsubsection{Timescale of years}

To estimate the annual variability of the blazar, we used data obtained between the years 2015 to 2024 in the $R$ and $V$ photometric bands, 2016 to 2024 in the $I$ band, and 2019 to 2024 in the $B$ band.
Part of the sample data had been previously reduced using different control and comparison objects; we thus analyse them separately. These data were obtained on the night of Jul 27, 1997, and during the period from Aug 11, 2016 to Jul 27, 2017.

\begin{table}
\centering
\caption{Results of the variability tests on yearly timescales. The first column shows the analysed time interval, the second lists the photometric bands, and the third indicates the number of observational nights. The next column shows the number of data points. Following that, we present the results of the $C_\Gamma$ and $F_\Gamma$ tests (with results in bold indicating variability). Finally, the last column reports the value of $\sigma_2$ as a measure of the observational error.}
{%
\begin{tabular}{ c  c  r r  r  r c }
\hline
\toprule
 Interval & Band & \multicolumn{1}{c}{Nights} & \multicolumn{1}{c}{$n$} & \multicolumn{1}{c}{$C_\Gamma$} & \multicolumn{1}{c}{$F_\Gamma$} & $\sigma_2$ \\
\midrule
2019--2024 & $B$ & 33 & 115 & \textbf{7.86} & \textbf{61.70} & 0.035 \\
2015--2024     & $V$ & 74 & 1554 & \textbf{8.40} & \textbf{70.52} & 0.018 \\
2015-2024    & $R$ & 77 & 3837 & \textbf{5.85} & \textbf{34.27} & 0.029 \\ 
2017--2024     & $I$ & 46 & 257 & \textbf{7.15}& \textbf{51.17} & 0.028 \\
\bottomrule
\end{tabular}
}
\label{tab:teststotal}
\end{table}

Results from the tests are shown in Table~\ref{tab:teststotal}. In this case, the results of both tests detected variability in all bands, confirming the strongly significant variability of this object on this timescale.

\subsubsection{Timescale of months}

We analysed the variability on the scale of months after grouping the differential photometry data by year of observation. In each case, the observations were never earlier than June or later than November; that is, the longest periods we considered were never longer than 6 months. To analyse the variability of the blazar on these scales, we took those years with at least more than one night of observation. Therefore, for this analysis we did not consider the year 1997, with only one observation night.

\begin{table}
\centering
\caption{Results of the variability tests on monthly timescales. Columns are the same as in Table \ref{tab:teststotal}. Results in bold indicate positive variability.}
\scalebox{0.85}{%
\begin{tabular}{ c  c  r r  r  r c }
\hline
\toprule
 Interval & Band & \multicolumn{1}{c}{Nights} & \multicolumn{1}{c}{$n$} & \multicolumn{1}{c}{$C_\Gamma$} & \multicolumn{1}{c}{$F_\Gamma$} & $\sigma_2$ \\
\midrule
 Aug 18 -- Sep 09, 2015  & $V$ & 2 & 29 & \textbf{5.93} & \textbf{35.12} & 0.024 \\
 & $R$ & 2 & 31 & \textbf{6.81} & \textbf{46.32} & 0.021 \\
 
 Jul 28 -- Sep 08, 2017 & $V$ & 11 & 101 & \textbf{3.37} & \textbf{11.38} & 0.030 \\
  & $R$ & 11 & 111 & \textbf{4.12} & \textbf{16.96} & 0.029\\
 & $I$ & 10 & 113 & \textbf{5.04}& \textbf{25.36} & 0.025 \\

Jun 07 -- Sep 03, 2019 & $B$ & 5 & 9 & 1.80 & 3.24 & 0.061 \\
 & $V$ & 16 & 260 & \textbf{7.61} & \textbf{57.97} & 0.010 \\
  & $R$ & 17 & 835 & \textbf{3.83} & \textbf{14.70} & 0.018 \\
 & $I$ & 5 & 9 & 0.58 & 2.95 & 0.172 \\

Aug 13 -- Sep 03, 2021 & $B$ & 2 & 7 & \textbf{5.41} & \textbf{29.22} & 0.005 \\
 & $V$ & 2 & 36 & 2.50 & \textbf{6.23} & 0.004 \\
  & $R$ & 2 & 36 & 1.62 & \textbf{2.64} & 0.004 \\
 & $I$ & 2 & 7 & \textbf{2.74} & 7.49 & 0.003 \\

Sep 16 -- Oct 24, 2022 & $R$ & 2 & 31 & \textbf{7.12} & \textbf{50.63} & 0.004 \\

Jul 11 -- Sep 19, 2023 & $B$ & 14 & 63 & \textbf{11.35} & \textbf{128.78} & 0.025 \\
 & $V$ & 26 & 527 & \textbf{17.47} & \textbf{305.06} & 0.012 \\
  & $R$ & 26 & 2230 & \textbf{14.40} & \textbf{207.44} & 0.014 \\
 & $I$ & 15 & 65 & \textbf{8.24} & \textbf{67.95} & 0.026 \\
Jun 18 -- Nov 11, 2024 & $B$ & 11 & 36 & \textbf{3.31} & \textbf{10.93} & 0.024 \\
     & $V$ & 16 & 580 & \textbf{4.97} & \textbf{24.68} & 0.019 \\
     & $R$ & 16 & 571 & \textbf{5.29} & \textbf{28.03} & 0.018 \\ 
     & $I$ & 12 & 39 & \textbf{2.76 }& \textbf{7.61} & 0.036 \\

\bottomrule
\end{tabular}
}
\label{tab:testsmensual}
\end{table}

Table \ref{tab:testsmensual} shows that, in those cases where both the $F_\Gamma$ and $C_\Gamma$ tests could not reject the null hypothesis, the number of photometric points was low (fewer than 10 per period considered). In contrast, in each case where at least 10 data points were available, the object was found to be variable according to at least one of the tests. Furthermore, the cases where no variability was detected by either test correspond to the highest values of photometric error ($\sigma_2$). This suggests that variations in photometric conditions may mask the intrinsic variability of the object.

\subsubsection{Interday scale}

In this case, we considered time intervals shorter than 10 consecutive days that included at least three nights of observations in the same photometric band. The results are presented in Table~\ref{tab:testsinterdia}. This table shows that all the periods considered turned out to be variable, with the notable exception of the one spanning the interval September 05 -- 13, 2017. We can note that during this period, the photometric error reaches its highest values, which could be affecting the results obtained.

\begin{table}
\centering
\caption{Results of the variability tests on interday timescales. Columns are the same as in Table \ref{tab:teststotal}. Results in bold indicate positive variability.}
\scalebox{0.85}{%
\begin{tabular}{ c  c  c  r  r  r  c}
\hline
\toprule
 Interval & Band & \multicolumn{1}{c}{Nights} & \multicolumn{1}{c}{$n$} & \multicolumn{1}{c}{$C_\Gamma$} & \multicolumn{1}{c}{$F_\Gamma$} & $\sigma_2$ \\
\midrule
 Jul 28 -- 31, 2017  & $V$ & 3 & 18 & \textbf{2.63} & \textbf{6.92} & 0.017 \\
 & $R$ & 3 & 25 & \textbf{5.63} & \textbf{31.69} & 0.004 \\

 Aug 10 -- 19, 2017  & $V$ & 4 & 37 & \textbf{3.39} & \textbf{11.48} & 0.023 \\
 & $R$ & 4 & 37 & \textbf{3.37} & \textbf{11.37} & 0.020 \\

 Sep 05 -- 13, 2017  & $V$ & 4 & 46 & 0.97 & 1.07 & 0.037 \\
 & $R$ & 4 & 49 & 1.14 & 1.29 & 0.040 \\

 Jul 18 -- 20, 2019 & $R$ & 3 & 233 & 2.52 & \textbf{6.34} & 0.028 \\
% & V & 3 & 4 & \textbf{6.96} & 48.48\\

 Aug 17 -- 26, 2019  & $V$ & 5 & 143 & \textbf{7.35} & \textbf{54.02} & 0.009 \\
 & $R$ & 5 & 139 & \textbf{7.30} & \textbf{53.33} & 0.009 \\

 Aug 31 -- Sep 03, 2019  & $V$ & 4 & 97 & \textbf{6.97} & \textbf{48.51} & 0.008 \\
 & $R$ & 4 & 90 & \textbf{10.15} & \textbf{103.00} & 0.005 \\

Jun 18 -- 23, 2023 & $V$ & 3 & 57 & \textbf{3.97} & \textbf{15.73} & 0.005\\
        & $R$ & 3 & 51 & \textbf{3.39} & \textbf{11.49} & 0.006 \\

Jul 08 -- 14, 2023 & $V$ & 7 & 40 & \textbf{2.63} & \textbf{6.91} & 0.019 \\
        & $R$ & 7 & 303 & \textbf{3.35} & \textbf{11.20} & 0.012 \\

  Aug 16 -- 18, 2023  & $V$ & 3 & 125 & \textbf{3.05} & \textbf{9.33} & 0.011 \\
  & $R$ & 3 & 126 & \textbf{3.17} & \textbf{10.03} & 0.010 \\

  Sep 13 -- 19, 2023 & $V$ & 6 & 287 & \textbf{5.91} & \textbf{34.94} & 0.007 \\
  & $R$ & 6 & 285 & \textbf{6.13} & \textbf{37.63} & 0.006 \\

  Sep 24 -- Oct 01, 2023 & $V$ & 4 & 7 & \textbf{5.91} & \textbf{34.94} & 0.008 \\
            & $R$ & 4 & 988 & \textbf{6.13} & \textbf{37.63} & 0.007 \\
Sep 06 -- 10, 2024 & $V$ & 5 & 283 & \textbf{5.04} & \textbf{25.36} & 0.006 \\
        & $R$ & 5 & 277 & \textbf{5.10} & \textbf{26.01} & 0.006 \\
Oct 29 -- Nov 02, 2024 & $V$ & 5 & 137 & \textbf{19.92} & \textbf{397.05} & 0.006 \\
        & $R$ & 5 & 145 & \textbf{8.83} & \textbf{78.03} & 0.010 \\
\bottomrule
\end{tabular}
}
\label{tab:testsinterdia}
\end{table}

 We also present the results for those nights for which photometry values were only available for a different set of \textbf{C} and \textbf{K} objects than those selected for the main study. These results can be seen in Table \ref{tab:interdiacell}. These nights were characterized by a high photometric error. We observe that the $F_\Gamma$ results always indicate variability, whereas this is not the case for the $C_\Gamma$ test. This strongly supports the notion that the $F_\Gamma$ test is more prone to yield false positives when dealing with noisy light curves, in contrast to $C_\Gamma$ \citep[see][]{zibecchi2017,zibecchi2020}.

\begin{table}
\centering
\caption{Results of the weighted $C_\Gamma$ and $F_\Gamma$ variability tests on interday timescales, performed for intervals with different control and comparison objects between 2016 and 2017. Columns are the same as in Table \ref{tab:teststotal}. Results in bold indicate positive variability.}
\scalebox{0.9}{%
\begin{tabular}{c c c r r r c}
\hline
\toprule
 Interval & Band & \multicolumn{1}{c}{Nights} & \multicolumn{1}{c}{$n$} & \multicolumn{1}{c}{$C_\Gamma$} & \multicolumn{1}{c}{$F_\Gamma$} & $\sigma_2$ \\
\midrule
 Aug 11 -- 27, 2016 & $B$ & 3 & 133 & 1.53 & \textbf{2.34} & 0.038\\
  & $V$ & 3 & 145 & 2.23 & \textbf{4.95} & 0.033 \\
   & $R$ & 4 & 139 & 1.94 & \textbf{3.78} & 0.034 \\
 & $I$ & 3 & 130 & 2.31 & \textbf{5.32} & 0.022 \\

 Oct 13 -- 26, 2016  & $R$ & 6 & 19 & \textbf{2.76} & \textbf{7.63} & 0.063 \\
 & $I$ & 6 & 12 & \textbf{8.56} & \textbf{73.28} & 0.017 \\

 Jul 18 -- 27, 2017  & $V$ & 4 & 42 & \textbf{7.8} & \textbf{60.88} & 0.014 \\
  & $R$ & 5 & 69 & 2.51 & \textbf{6.30} & 0.034 \\
 & $I$ & 4 & 34 & \textbf{3.67} & \textbf{13.46}  & 0.030 \\

\bottomrule
\end{tabular}
}
\label{tab:interdiacell}
\end{table}

\subsubsection{Microvariability}

In this section, we describe the results of the statistical analysis performed on the data from all nights, considering each night individually and across all photometric bands (see Table~\ref{tab:intra}; only the first three nights are shown here, and the full table is available as supplementary material). In some cases, the low number of data available per night (less than 5 data points) caused the test results not to be statistically significant. Although we obtained the $C_\Gamma$ and $F_\Gamma$ parameters for these cases, the results for these nights are not shown in the tables. The results obtained in this time frame were mostly negative. However, on nights where both tests were positive for the $R$ and $V$ bands, and there were sufficient data, the intervals were subdivided. The cases in which we were able to perform this analysis correspond to the new observations conducted between 2023 and 2024. This allowed for the detection of microvariability on a timescale of 5 and 3 hours.

\begin{table}

\caption{Microvariability test results. Columns show the night, band, number of data points, $C_\Gamma$ and $F_\Gamma$ test results (bold indicates variability), and $\sigma_2$ as the observational error estimate. The complete version of this table is available in the online supplementary material.} 
\centering
\begin{tabular}{c c c c c c |}
\hline
\toprule
Date & Band &  \multicolumn{1}{c}{$n$} & \multicolumn{1}{c}{$C_\Gamma$} & \multicolumn{1}{c}{$F_\Gamma$} & $\sigma_2$ \\
\midrule
Jul 27, 1997 & $V$ & 69 & 1.04 & 1.08 & 0.007  \\
Aug 13, 2015 & $V$ & 19 & 0.91 & 1.22 & 0.015  \\
             & $R$ & 19 & 0.94 & 1.13 & 0.015  \\    
Sep 15, 2015 & $V$ & \phantom{1}9 & 1.28 & 1.64 & 0.012 \\  
& $R$ & \phantom{1}9 & 0.99 & 1.01 & 0.009 \\
Aug 11, 2016 & $B$ & 31 & 1.32 & 1.49 & 0.005 \\
         & $V$ & 31 & 1.28 & 1.63 & 0.004  \\
         & $R$ & 31 & 0.99 & 1.02 & 0.005  \\
         & $I$ & 31 & 1.07 & 1.15 & 0.004  \\
%  ...    & ... & ... & ... & ... & ... \\
%Nov 11, 2024 & $V$ & \phantom{1}5 & 1.01 & 1.02 & 0.005 \\
%             & $R$ & \phantom{1}5 & \textbf{2.79} & 7.76 & 0.002 \\       
         
\bottomrule

\end{tabular}
\label{tab:intra}

\end{table}

\subsection{Light curves in the standard system}

We provide an overview of selected light curves calibrated to the standard system, using the standard magnitudes of the comparison stars published by \cite{hamuy} for the transformation.

\begin{figure}
    \centering
    \includegraphics[width=.5\textwidth]{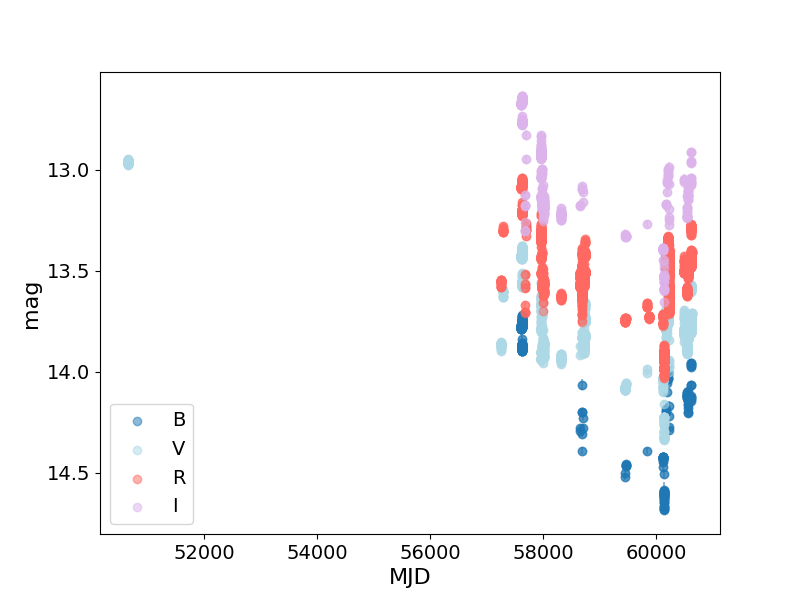}
    \caption{Standard magnitude light curves of PKS\,2155$-$304 in the $B$, $V$, $R$, and $I$ bands from the entire observation campaign.}
    \label{fig:1997-2024}
\end{figure}

In Figure \ref{fig:1997-2024} we show the complete observation campaign, where it is clear that the optical bands are strongly correlated. Noticeably, a peak in the $V$ band is observed in the 1997 data, reaching a magnitude of 12.95, a brightness level that is never reached again during the rest of the campaign reported here.

\begin{figure}
    \centering
    \includegraphics[width=.5\textwidth]{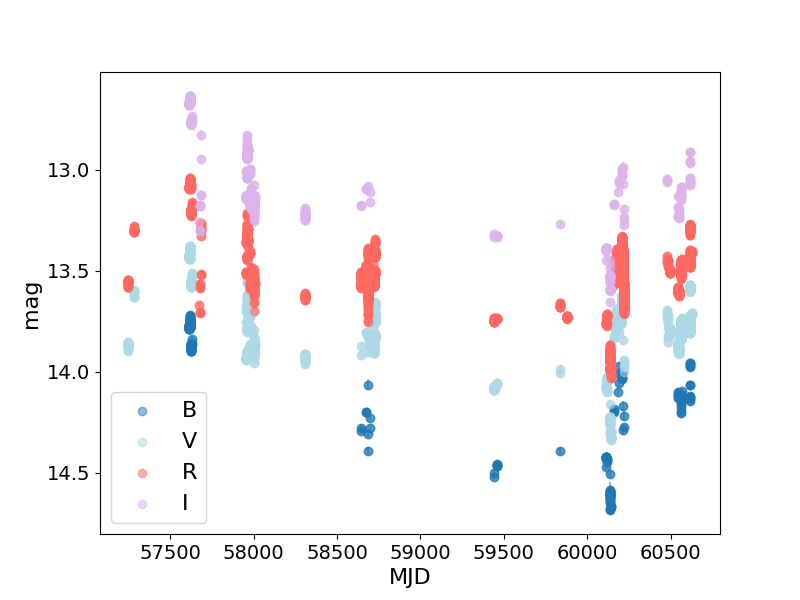}
    \caption{Standard magnitude light curves of PKS\,2155$-$304 in the $B$, $V$, $R$, and $I$ bands from the year 2015 to 2024.}
    \label{fig:2015-2024}
\end{figure}

We illustrate the light curves from 2015 to 2024 in Figure \ref{fig:2015-2024},  the period with the highest density of data points over time. The mean magnitude values of the object during this period were $B$ = 14.04, $V$ = 13.77, $R$ = 13.54, and $I$ = 13.03. Maximum activity levels, corresponding to the highest fluxes observed during the monitoring period, were observed in all bands on August 19, 2016 (MJD = 57614). At that time, the magnitude values were $B$ = 13.72, $V$ = 13.38, $R$ = 13.04, and $I$ = 12.64. Conversely, the lowest activity levels that is, when the source flux reached its minimum, were recorded in July 2023, with magnitude values of $B$ = 14.68, $V$ = 14.34, $R$ = 14.03, and $I$ = 13.65 ($\Delta B = 0.96$, $\Delta V = 0.96$, $\Delta R = 0.99$ and $\Delta I = 1.01$). These minimum activity values were reached just a few days before the predicted flare (see section \ref{sec:flare}).

\begin{figure}
    \centering
    \includegraphics[width=.5\textwidth]{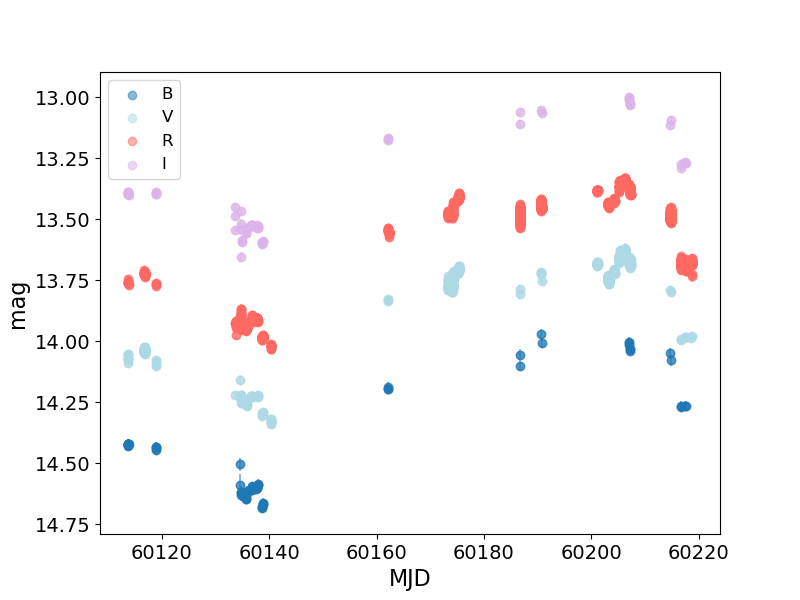}
    \caption{Standard magnitude light curves of PKS\,2155$-$304 in the $B$, $V$, $R$, and $I$ bands corresponding to the year 2023.}
    \label{fig:2023}
\end{figure}

\begin{figure}
    \centering
    \includegraphics[width=.5\textwidth]{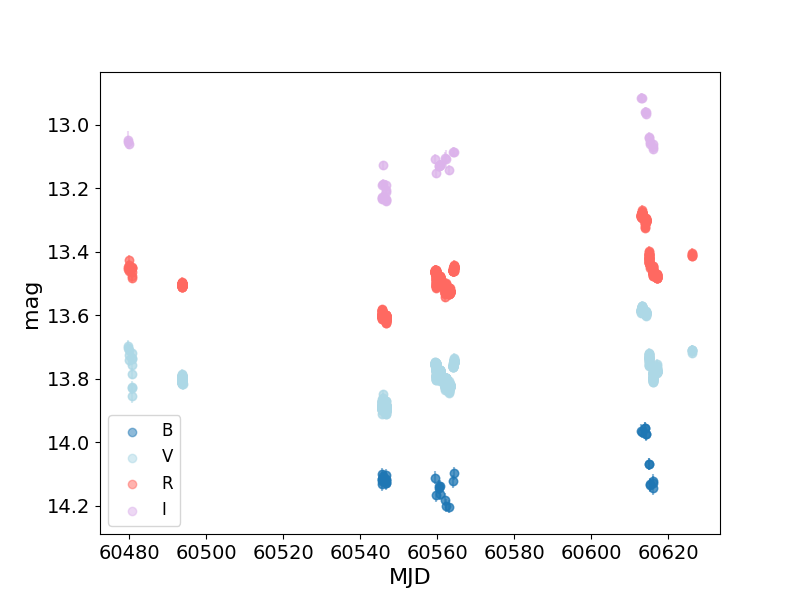}
    \caption{Standard magnitude light curves of PKS\,2155$-$304 in the $B$, $V$, $R$, and $I$ bands corresponding to the year 2024.}
    \label{fig:2024}
\end{figure}

In Figures \ref{fig:2023} and \ref{fig:2024}, we show the light curves corresponding to the years 2023 and 2024, respectively. During these intervals, we have several observations in the four bands, and these were the years where we achieved the best temporal coverage of the object. Furthermore, the above mentioned microvariability events on timescales shorter than 5 hours were detected during these years.

In the year 2023, we observed magnitude variations in the $B$, $V$, $R$, and $I$ bands of 0.70, 0.72, 0.72 and 0.66 respectively. The average magnitudes during this period in each band were 14.43, 13.79, 13.59 and 13.39. In contrast, in the year 2024, we found variations of 0.24, 0.34, 0.35 and 0.33 in magnitude, and mean values of 14.10, 13.78, 13.48 and 13.11. We can conclude, then, that the source was more active in 2023.

\begin{figure}
    \centering
    \includegraphics[width=.5\textwidth]{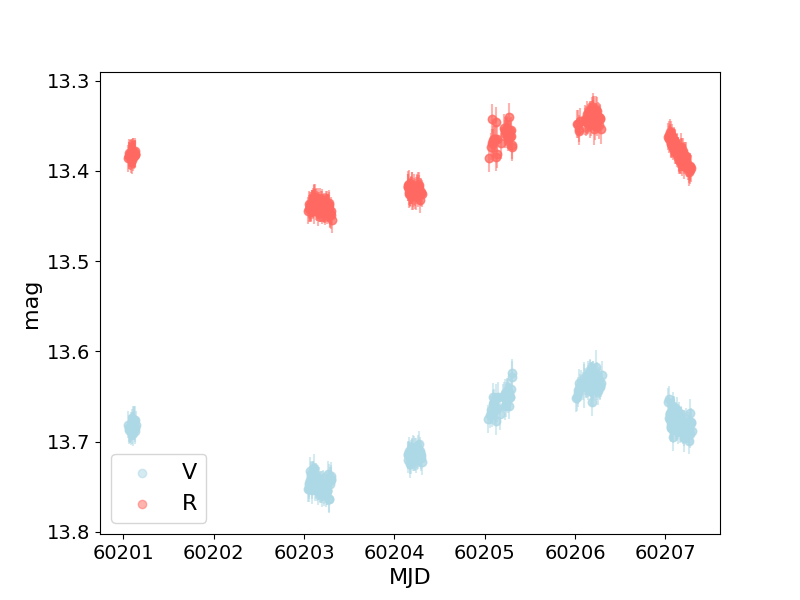}
    \caption{Standard magnitude light curves of PKS\,2155$-$304 in the $V$ and $R$ bands corresponding to the interval Sep 13 -- Sep 19, 2023}
    \label{fig:2023-09}
\end{figure}

Figure \ref{fig:2023-09}, shows the light curve for the interval September 13--19, 2023. During this period, we were able to achieve a rather dense temporal sampling of the blazar's flux, thus highlighting the peculiar variability behaviour of the source, with the data points appearing to follow a sinusoidal-like pattern. We measured magnitude variations of 0.14 and 0.12 in the $V$ and $R$ bands, respectively, with values ranging from 13.76 to 13.62 in $V$ and from 13.45 to 13.33 in $R$.

\begin{figure}
    \centering
    \includegraphics[width=.5\textwidth]{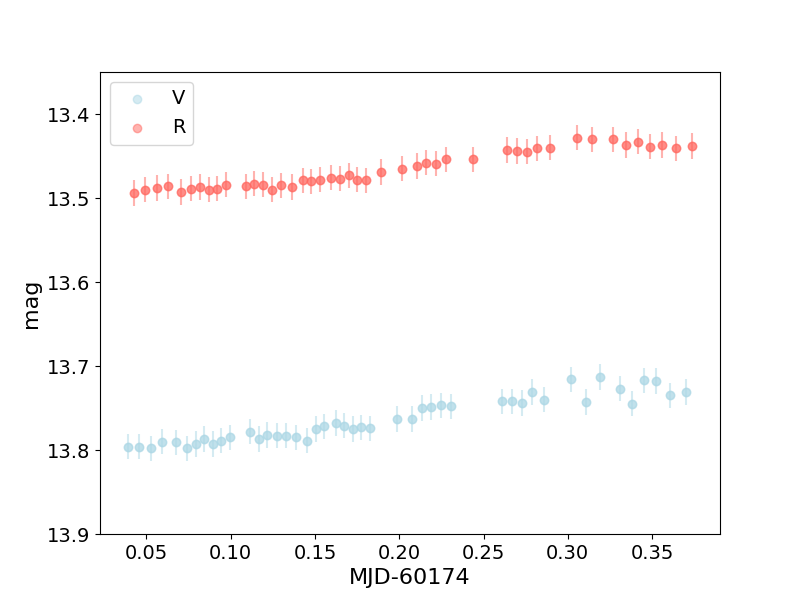}
    \caption{Standard magnitude light curves of PKS\,2155$-$304 in the $V$ and $R$ bands corresponding to Aug 17, 2023}
    \label{fig:2023-08-17}
\end{figure}

\begin{figure}
    \centering
    \includegraphics[width=.5\textwidth]{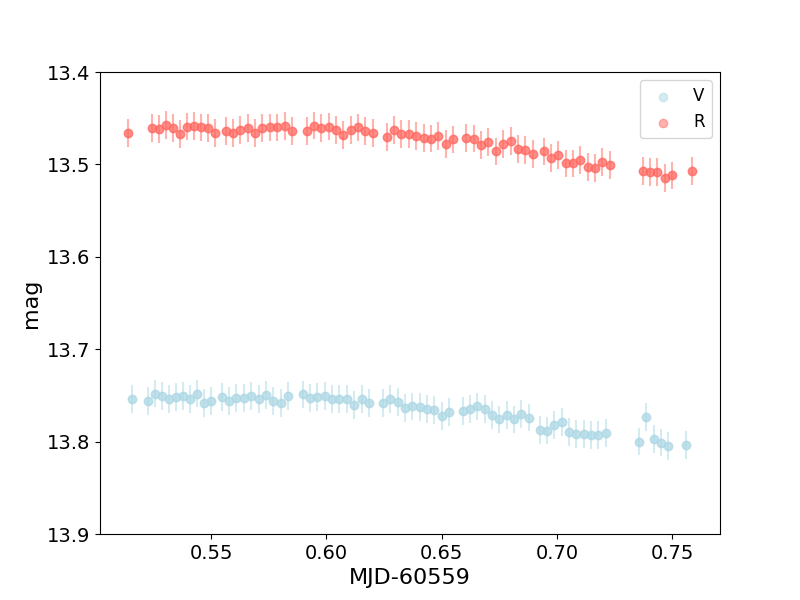}
    \caption{Standard magnitude light curves of PKS\,2155$-$304 in the $V$ and $R$ bands corresponding to Sep 06, 2024}
    \label{fig:2024-09-06}
\end{figure}

Finally, in Figures \ref{fig:2023-08-17} and \ref{fig:2024-09-06}, we present the light curves corresponding to August 17, 2023, and September 06, 2024. These are examples of nights where variability was detected in both the $V$ and $R$ bands, with the observation on August 17 showing variability within a 5 hour interval, with total magnitude variations of 0.09 and 0.06 in $V$ and $R$, respectively. Similarly, the observation on September 06, 2024 exhibited variability within a 3-hour interval, with a total magnitude variation of 0.05 in both $V$ and $R$ bands.



\subsection{X-ray correlation analysis}

We compared the optical flux in the $V$ and $R$ bands with the $X$-ray counts (\textit{soft}, \textit{medium}, and \textit{hard}). These two optical bands were selected as they provided the largest amount of data and the most extensive temporal coverage. We computed the ZDCF for different time lags ($Z$ and $\tau$, respectively), where $Z$ quantifies the significance of the correlation between the bands.

We conducted two different approaches. In the first one, no constraints were imposed on the time lag. This approach yielded highly significant correlations at time lags close to the dataset's sampling interval, which led us to discard these results. In the second one, we restricted the time lag up to 100 days, a threshold within which time delays are expected to remain physically meaningful in the context of blazar emission models. Larger lags are more likely to reflect unrelated variability patterns rather than genuine correlations between the X-ray and optical bands \citep{maoz}. 

\begin{figure}
    \centering
    \includegraphics[width=.5\textwidth]{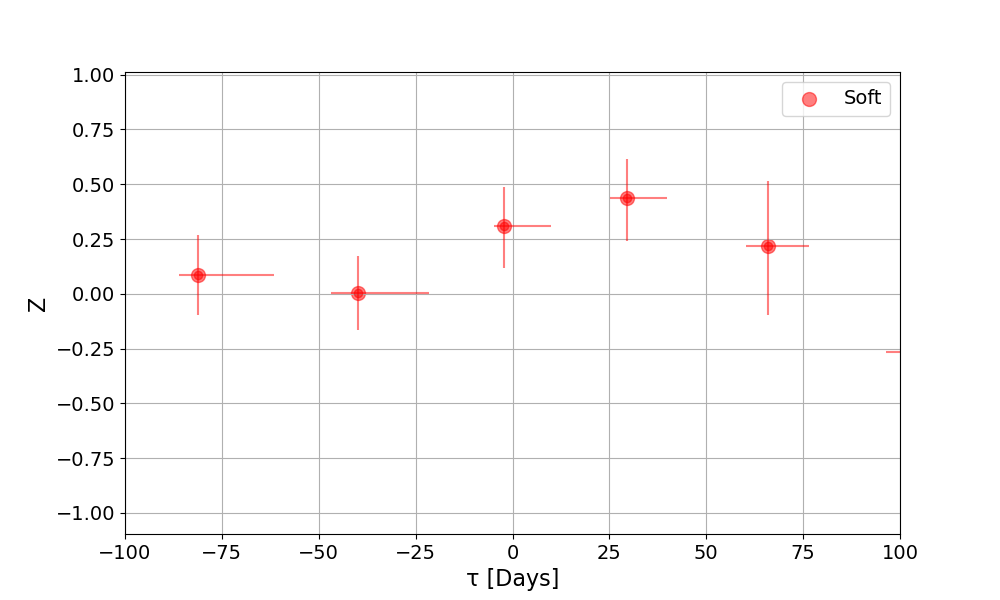}
    \caption{ZDCF results between the $R$ bands and the soft $X$-Ray band for different time lags $\tau$.}
    \label{fig:CorrelacionR}
\end{figure}

In this case, the results showed weak correlations with large associated uncertainties, with no significant differences between bands or fluxes. The significance values of $Z$ never exceeded 0.5 and, in all cases, exhibited high dispersion in their associated errors. If a $Z$ value were statistically significant, we would expect a clustering of $Z$ results with lower significance around this point; however, this does not occur for any data point. As an example, Figure \ref{fig:CorrelacionR} illustrates the correlation between the \textit{soft} band and $R$.

Due to the large dispersions, it is concluded that the observed correlation is weak, a result consistent with that found by \cite{treves}. However, this differs from the findings of \cite{urry1998}, where they report a strong correlation between X-rays and the optical band within a time lag of less than 3 hours. In that case, data in both the X and optical bands were acquired simultaneously. The weak correlation we found here does not support the existence of certain long-term physical processes in the AGN, given that the observed variations are chromatic. For example, it rules out the possibility that the jet is precessing, contrary to the results obtained by \cite{zheng}. This is because a change in the jet orientation with respect to the line of sight is a geometric effect that would be expected to produce achromatic variability across the entire electromagnetic spectrum due to changes in the Doppler boosting factor (see, e.g., \citealp[]{raiteri}).

\subsection{Flare}\label{sec:flare}

Based on a personal communication (B. Kapanadze, 2023), we were informed that, in the time interval between July and August  2023, the blazar should experience a sustained significant increase in its flux. His prediction was based on Swift-XRT observations that detected an ongoing X-ray flare with extreme intraday variability. According to current physical models, such as the one-zone Synchrotron Self-Compton scenario, this X-ray activity indicated an expected enhancement in multiwavelength emission. \cite{Kapanadze}.

\begin{figure}
    \centering
    \includegraphics[width=.5\textwidth]{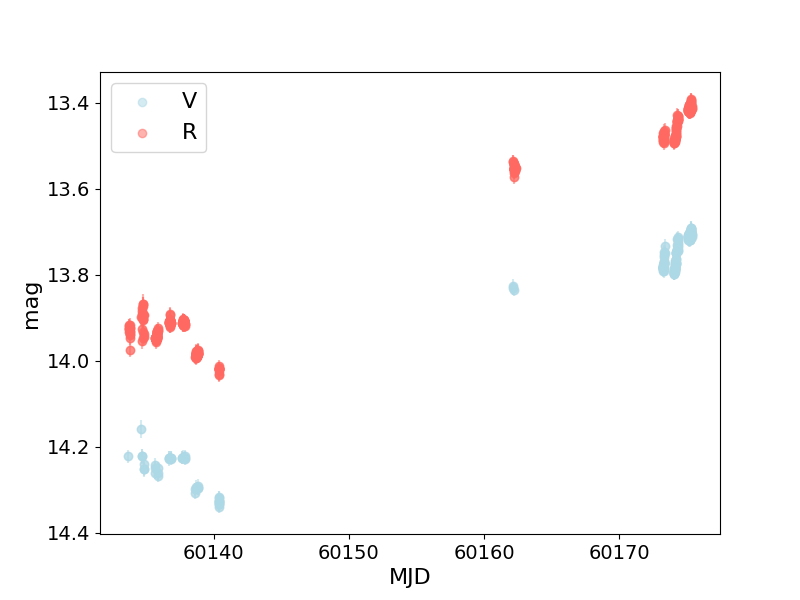}
    \caption{Standard magnitude light curves of PKS\,2155$-$304 in the $V$ and $R$ bands corresponding to Jul 08 -- Aug 18, 2023. A sustained increase in the measured fluxes is observed.}
    \label{fig:flare}
\end{figure}

The new observations conducted by the group reported a significant increase in the object's luminosity, reflected in variations in its magnitude across the photometric bands, during the period between the nights of July 08 and August 18, 2023 (MJD 60136–60174; see Figure \ref{fig:flare}). The results of the statistical tests applied to this period are shown in Table \ref{tab:flare}.  In the two photometric bands where observations were made ($R$ and $V$), both tests were clearly positive, confirming the object's variability. The significant increase of approximately half a magnitude ($\Delta R = 0.47$; $\Delta V = 0.50$), observed between July 14 and August 5 (MJD 60139-60161), provides solid evidence of the occurrence of the flare, with these changes in $R$ and $V$ magnitudes corresponding to a flux increase of 54\% and 58\%, respectively. Subsequent measurements taken on August 18 showed that this change further increased to $\Delta R = 0.61$ and $\Delta V = 0.62$, corresponding to a total flux increase of 75\% in $R$ and 77\% in $V$. It is worth noting that it was during this period that we managed to measure microvariability over 5 hours, corresponding to the night of August 17. Altogether, these results provide strong evidence supporting the existence of the predicted flare, further highlighting the object's intrinsic variability.

\begin{table}
\centering
\caption{Results of the $C_\Gamma$ and $F_\Gamma$ variability tests conducted between Jul 08 -- Aug 18, 2023. The first column shows the analysed time interval, the second lists the photometric bands, and the third indicates the number of observational nights. The next column shows the number of data points. Following that, we present the results of the $C_\Gamma$ and $F_\Gamma$ tests (with results in bold indicating variability). Finally, the last column reports the value of $\sigma_2$ as a measure of the observational error.}
\scalebox{0.85}{%
\begin{tabular}{ c  c  r r  r  r c }
\hline
\toprule
 Interval & Band & \multicolumn{1}{c}{Nights} & \multicolumn{1}{c}{$n$} & \multicolumn{1}{c}{$C_\Gamma$} & \multicolumn{1}{c}{$F_\Gamma$} & $\sigma_2$ \\
\midrule
 Jul 08 -- Aug 18, 2023  & $V$ & 11 & 168 & \textbf{13.70} & \textbf{187.77} & 0.013 \\
 & $R$ & 11 & 456 & \textbf{19.38} & \textbf{375.47} & 0.007 \\
\bottomrule
\end{tabular}
}
\label{tab:flare}
\end{table}

\section{Data analysis}

\subsection{Spectral index behaviour}

We reconstructed the changes of the object's color over time by means of the measured magnitudes in the $R$ and $V$ bands. The corresponding data series starts in 2015, the year from which we have data in both the $R$ and $V$ bands. Since the optical spectrum follows a power-law distribution (in the form $F \propto \nu^{\alpha}$), the spectral index $\alpha$ can be calculated using %
 \begin{equation}
    \alpha = \frac{\log(F_{\lambda_2})-\log(F_{\lambda_1})}{\log(\lambda_1)-\log(\lambda_2)},
    \label{eq:IE}
\end{equation}
where fluxes are obtained from the standard magnitudes through the standardized zero-magnitude fluxes from \cite{bessel}. Given that the redshift is known, we applied a correction for the cosmological redshift:
\begin{equation}
    F=\frac{F^0}{(1+z)^2}\label{eq:cosm}.
\end{equation}
In this equation, $F$ represents the observed flux, while $F^0$ corresponds to the flux in the source's rest frame. These fluxes were computed in mJy.
No correction for Galactic extinction was applied; anyway, extinction in the direction of PKS\,2155$-$304 is low \citep[$E(B-V)=0.019$\,mag,][]{2011ApJ...737..103S}.

\begin{figure}
    \centering
    \includegraphics[width=.5\textwidth]{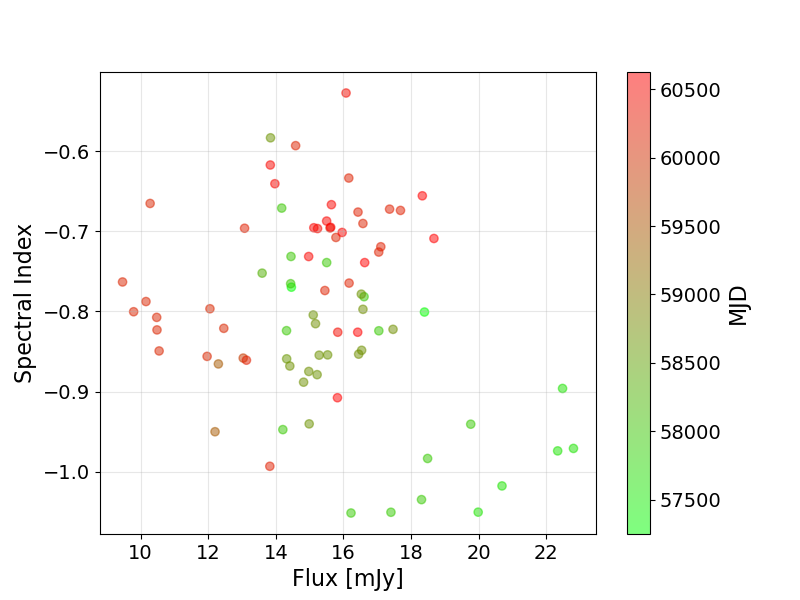}
    \caption{Spectral index versus flux for PKS\,2155$-$304. Each point represents a measurement of the spectral index and corresponding flux, with color indicating the MJD of the observation.}
    \label{fig:IC}
\end{figure}

Using the calculated $\alpha$ and the corresponding flux values, we constructed a spectral index versus flux diagram, incorporating the time variable to identify possible patterns (see Figure \ref{fig:IC}). Although a mild bluer-when-brighter behavior can be observed, the data dispersion is too large to draw any firm conclusions, and the fits performed were not statistically significant. The maximum value of $\alpha$ was $-0.53$ and occurred on August 18, 2024 (MJD 60540), while the minimum value was $-1.05$ on July 26, 2017 (MJD 57960), with a mean value of $-0.8$ with a standard deviation of 0.12.
 
%\textbf{We fitted a straight line using the Least-Mean-Squares method to relate the evolution of the spectral index with time,}

\begin{figure}
    \centering
    \includegraphics[width=.5\textwidth]{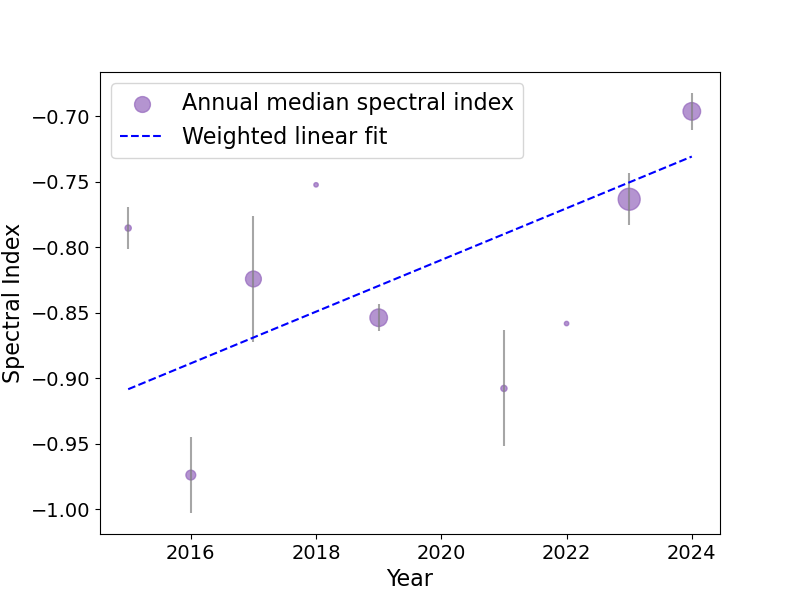}
    \caption{Temporal evolution of the spectral index of PKS\,2155$-$304. Point size is weighted by the number of observations. A weighted linear fit is included to highlight the long-term trend.}
    \label{fig:Ajuste}
\end{figure}

In order to study the long-term temporal behaviour of the spectral index, we performed an annual binning of the data and computed the median spectral index $\bar{\alpha}$ for each year, rather than fitting the full time series in MJD. This approach reduces the influence of outliers or extreme values, as the median is more robust than the mean. We then fitted a linear model to the yearly median values using a weighted least-squares method, where each year is weighted by its corresponding number of observations. The fit involves two free parameters, $a$ and $b$, with $t$ representing time in years:
\begin{equation}
    \bar{\alpha} = at+b.
\end{equation}
The resulting best-fit parameters were: $a = 1.97\times10^{-2}$ and $b = -40.69$
As shown in Figure~\ref{fig:Ajuste}, the overall trend indicates an increase in $\alpha$, suggesting a flattening of the spectrum over time.

%\textbf{with two free parameters, a and b. $t$ represents time in Modified Julian Date (MJD). The fit to our data yielded: $a=5.99\times10^{-5}$ and $b=-4.36$. As shown in Figure \ref{fig:Ajuste}, it can be observed that the overall trend of the index is increasing, indicating that the spectrum became flatter.}

%\begin{figure}[h!]
%    \centering
%    \includegraphics[width=.45\textwidth]{Ajuste.png}
%    \caption{Spectral Index vs. MJD}
%    \label{fig:Ajuste}
%\end{figure}

\subsection{Optical quasi-periodicity}

\begin{figure*}
    \centering
    \includegraphics[width=\textwidth]{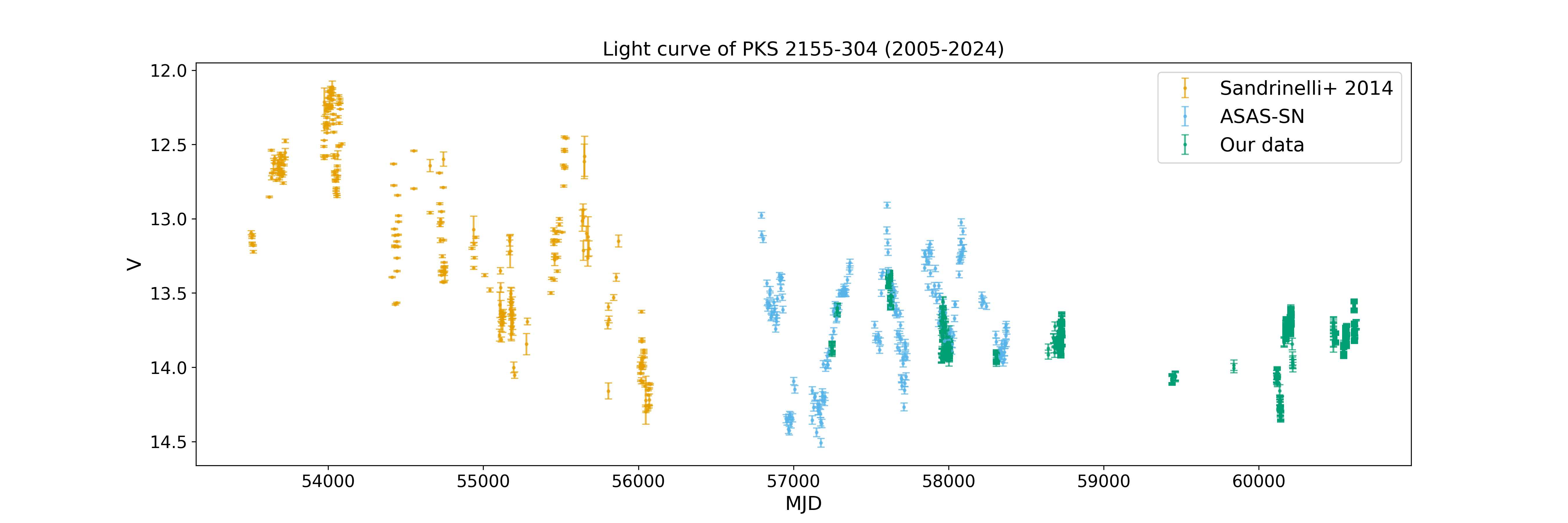}
    \caption{A historic light curve for PKS\,2155-304 in the $V$ band. The orange data correspond to \citet{sandrinelli14}, cyan to the ASAS-SN survey \citep{Kochanek17}, while our data are coloured green. Our data consistently match the ASAS-SN survey data when overlapped. The source is undergoing a relatively low-activity phase in comparison to the data published ten years before our first observing run.}
    \label{fig:lc_comparison_literature}
\end{figure*}

As stated in Sec. \ref{sec:Introduction}, several studies have found conflicting evidence of quasi-periodicities on the light curves of PKS\,2155$-$304. We thus aim at analysing our variability data to contribute, if possible, to the ongoing discussion. As a reference frame, we first compared our $V$ data together with the $V$ data from the All-Sky Automated Survey for Supernovae (ASAS$-$SN, \citealp{Kochanek17}), following \citet{zheng}, and the $V$ data from \citet{sandrinelli14} (see Figure \ref{fig:lc_comparison_literature}). We chose the $V$ band for this comparison since it is the only band in common between our data, the ASAS$-$SN data and the data published by \citet{sandrinelli14}. The first months of our data overlap with the last months from the ASAS$-$SN data. The magnitude we observed of PKS\,2155$-$304 is consistent with the magnitudes reported by ASAS-SN. It is evident that PKS\,2155$-$304 is now ongoing a relatively low activity phase, since the data from \citet{sandrinelli14} report consistently larger fluxes and variability amplitude.

Since the $R$ band has the largest number of observations within our set of data, it was selected as a reference for all the following analyses in this section; however, the results do not change significantly when other optical bands are used.

Given the fact that the full light curve (see Figure \ref{fig:1997-2024}) is densely sampled in intra-month scales but not in intra-year scales, we decided to look for quasi-periodicity only in intra-month scales (i.e., with periods of less than $\sim$ 30 days), to avoid artifacts and uncertainties due to the sparse population of data. The data were thus subdivided into nine segments, each separated by more than 100 days from the previous, and numbered accordingly from 1 to 9 in increasing order from oldest to newest. The choice of 100 days to separate the segments responds to the fact that the source is not observable from CASLEO for $\sim$100 days a year. Since the gaps are greater than the periods we are looking for, we focused only on each segment in particular.

We then implemented a Phase Dispersion Minimization (PDM) algorithm, following \cite{stellingwerf}. This approach does not rely on the functional form of the expected periodicity, and it also is less prone to distortions due to outliers or uneven sampling of the data. It consists on looking iteratively for the phase value that minimizes the dispersion of the data when folded into that phase. The statistic derived to measure the clusterisation of data points after being folded is $\theta$, and a lower value of $\theta$ indicates a smaller dispersion after folding. 

We find candidate periods for segments 5, 8 and 9 of 13.6, 29.7 and 19.6 days, respectively, with the PDM method. To evaluate the significance of these results we tested the null hypothesis through a surrogate test, by randomly shuffling the magnitude data to derive 1E4 surrogate light curves. From this test, a $p$-value is derived, corrected by the number of period trials and by the number of tested samples (i.e., segments). The $p$-values for these results are all $p<1E-4$. This would suggest, in principle, that the data do show structure when compared against random curves.

In all cases we checked for degeneration on the periodicity results. If gaps between observation shifts coincide with any given candidate period, it may hint at the algorithm finding spurious periodicity due to the sampling rate. To assess this, we calculated the $\Delta T$, in days, between consecutive observations for each segment. Most shifts were comprised of several consecutive nights, thus yielding $\Delta T<1$, but were discarded since periodicity analyses were performed for periods between 2 and 30 days. Each remaining $\Delta T$ and its frequency per segment are shown in Table~\ref{tab:deltaT}. We note that, for segment 5, several candidate periods are exact multiples or fractions of some sampling frequencies, since there are gaps between observations of 14.3 or 6.5 days. We note that there are no $\Delta T$ values close to the proposed periods of $\sim$19 or $\sim$29 days for segments 8 and 9, respectively.

\begin{table}
\caption{Values and occurrences of $\Delta T$ between consecutive observation nights, for all $\Delta T>1$ day, for each segment. In column 1 we list the segment ID, and in column 2 the value of $\Delta T$.}
\label{tab:deltaT}
\centering
\begin{tabular}{c c c c c c}
\hline\hline
Segment & $\Delta T$ & Segment & $\Delta T$ & Segment & $\Delta T$ \\ 
& [days] & & [days] & & [days]  \\
\hline
 5 & 33.4  & 8 & 21.8 & 9 & 51.8 \\
   & 21.7  &   & 14.3 &   & 48.7 \\
   & 14.3  &   & 11.3 &   & 12.8 \\
   & 6.5   &   & 10.9 &   & 12.7 \\
   & 4.8   &   & 10.3 &   & 8.9  \\
   &       &   & 7.9  &   &       \\
   &       &   & 3.8  &   &  \\
   &       &   & 2.9  &   &  \\
   &       &   & 2.0  &   &  \\
\hline
\end{tabular}
\end{table}

We cannot rule out, however, the possibility of any or all of these candidates being artifacts of our analysis. The periods we find do not coincide between segments, and in the case of segment 8, its candidate period would be detected over 3.5 cycles in our segmented data, which would be a marginal detection. Moreover, being a test on AGN data, comparing directly against random data might be unproductive, since AGN are known to display red noise \citep{press,vaughan}.

Thus, we proceeded to perform another surrogate test, this time by creating 1E4 synthetic, random light curves with red noise, following the method described by \citet{timmerkoenig}. We derived the power spectrum of our data for a given segment, and its spectral index. The same spectral index, $\alpha$, was used for the power law spectrum of the synthetic light curves, $P\propto f^{\alpha}$. The power spectra of the data yielded spectral indices $\alpha \sim 0.2\,-\,0.3$. Even considering min-max values, $\alpha$ remained always below 1, which is fairly flat. As before, we derive trials and sample corrected p$-$values for each segment.

\begin{figure}
    \centering
    \includegraphics[width=0.5\textwidth]{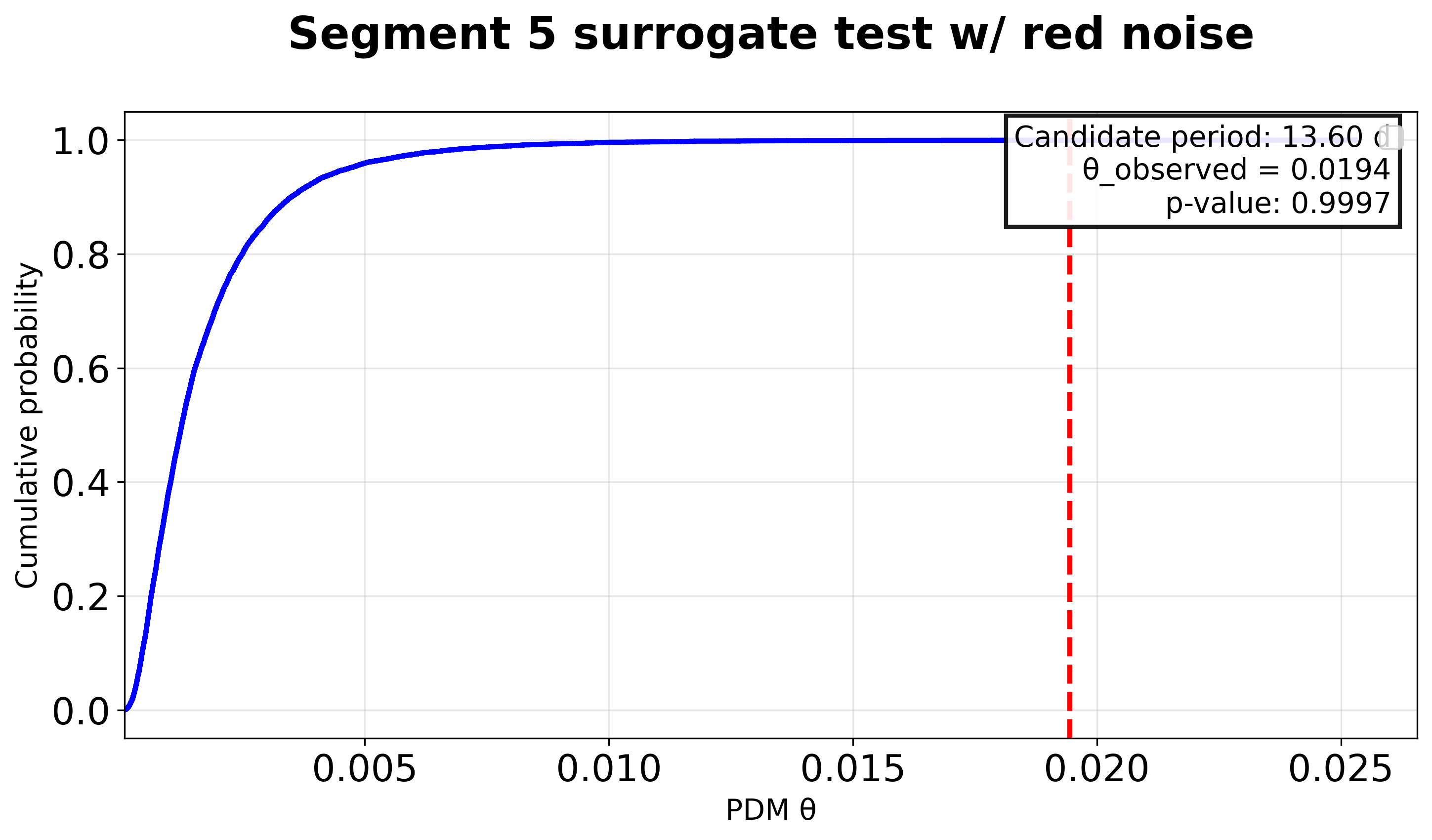}
    \caption{PDM $\theta$ statistic cumulative distribution function for the red noise surrogate test, for data segment 5. The red dashed line marks the $\theta$ value found for this segment, $\theta=0.0194$, which corresponds to a period of $\sim 13.6$ days. The derived $p$-value for this period is $p=0.9997$.}
    \label{fig:pdm_s5}
\end{figure}

\begin{figure}
    \centering
    \includegraphics[width=0.5\textwidth]{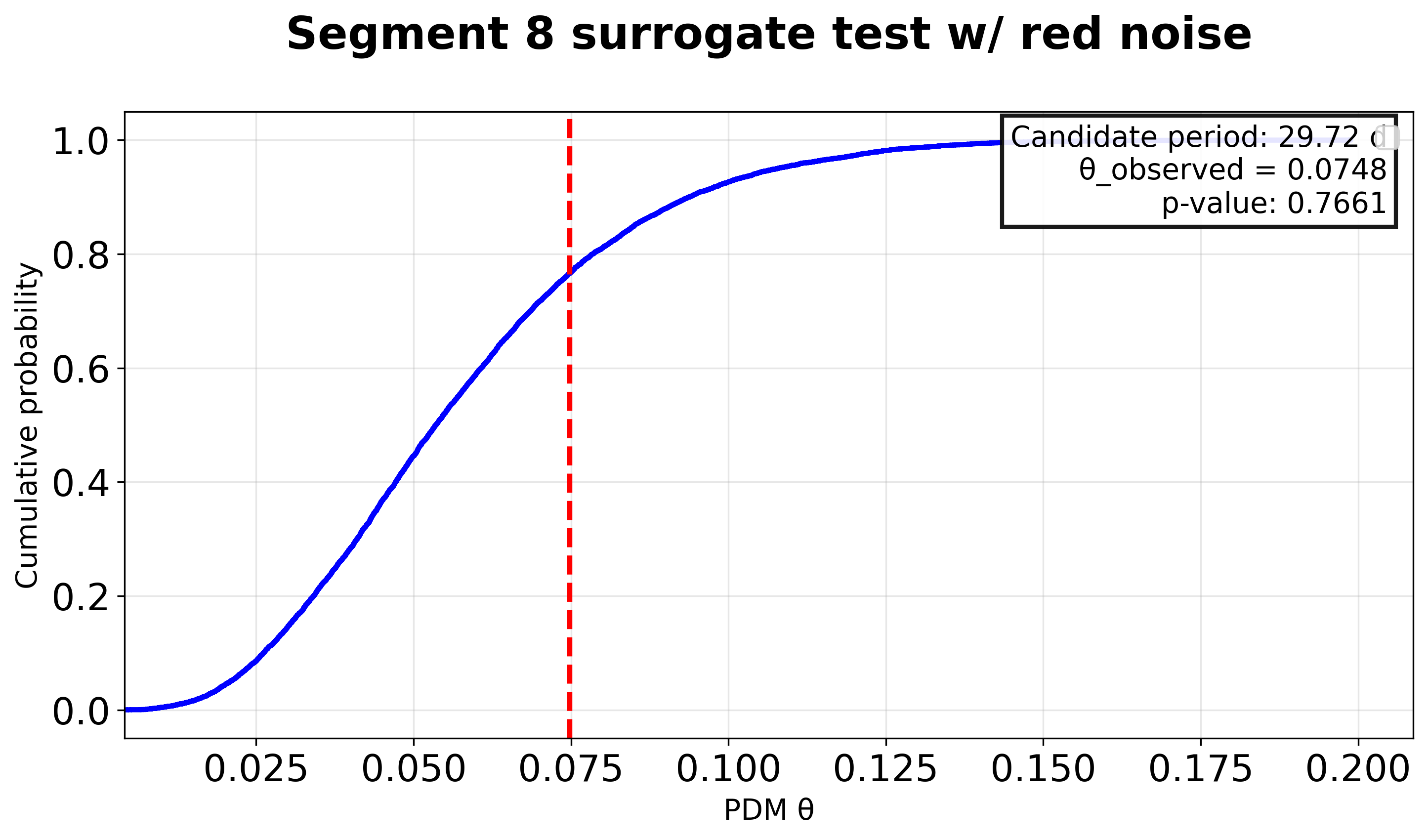}
    \caption{PDM $\theta$ statistic cumulative distribution function for the red noise surrogate test, for data segment 8. The red dashed line marks the $\theta$ value found for this segment, $\theta=0.0748$, which corresponds to a period of $\sim 29.7$ days. The derived $p$-value for this period is $p=0.7661$.}
    \label{fig:pdm_s8}
\end{figure}

\begin{figure}
    \centering
    \includegraphics[width=0.5\textwidth]{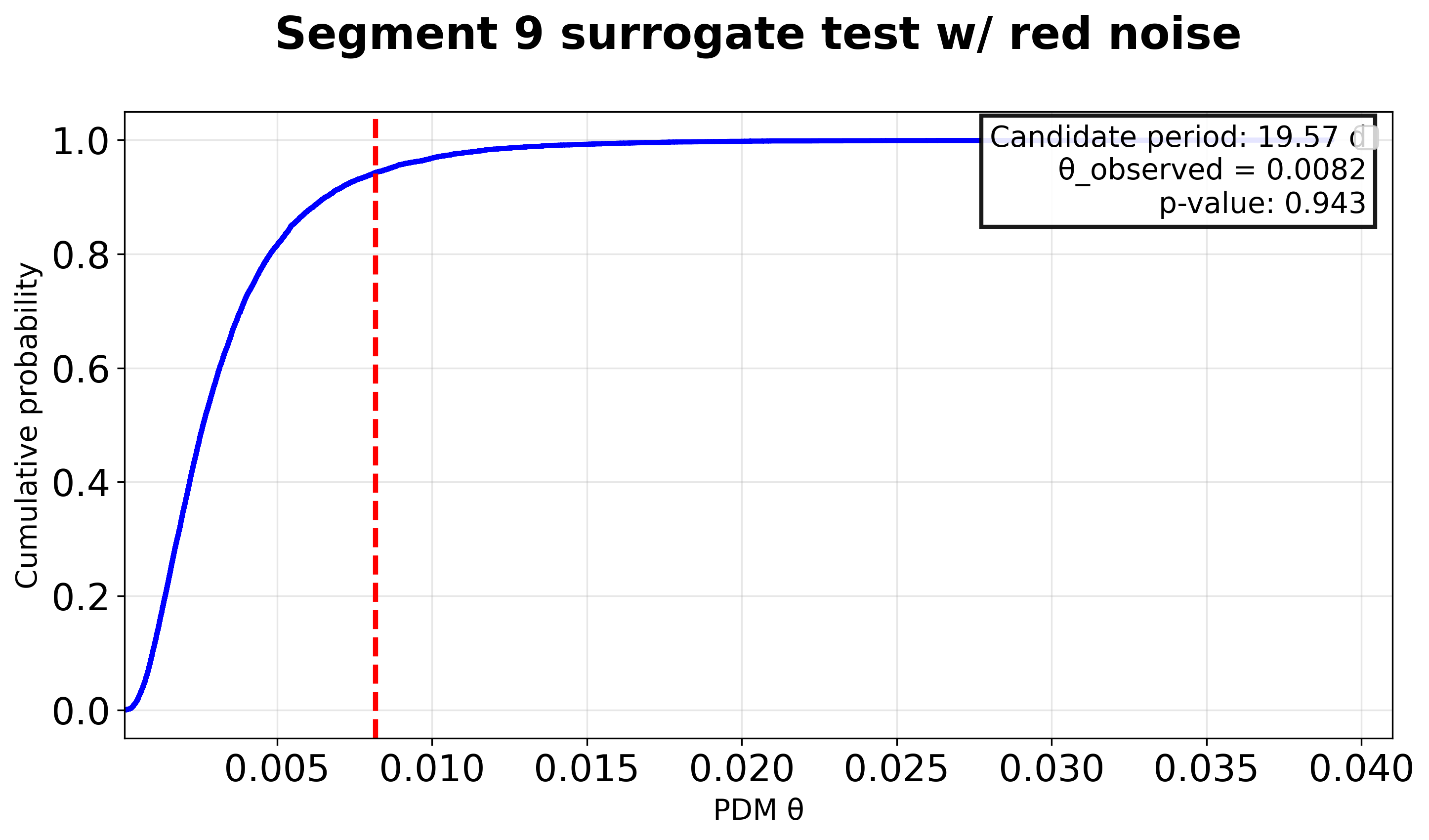}
    \caption{PDM $\theta$ statistic cumulative distribution function for the red noise surrogate test, for data segment 9. The red dashed line marks the $\theta$ value found for this segment, $\theta=0.0082$, which corresponds to a period of $\sim 19.6$ days. The derived $p$-value for this period is $p=0.9430$.}
    \label{fig:pdm_s9}
\end{figure}

We show the cumulative distribution function of the red-noise surrogate tests for each derived $\theta$ (and their corresponding periods) in Figures \ref{fig:pdm_s5}, \ref{fig:pdm_s8} and \ref{fig:pdm_s9}, respectively. The cumulative distribution function is a simple way to visualize how many occurrences of this $\theta$ value were found in the surrogate light curves, i.e. to visualize the derived global $p$-value. This test yielded $p$-values of 0.9997, 0.7661 and 0.9430, for segments 5, 8 and 9, respectively, which rule out the presence of any significant periodicity. We note that the power spectrum used for the synthetic light curves was not strongly red ($\alpha\sim0.3$), and thus if a stronger red noise component was present in the synthetic curves the significance would drop even further. We thus conclude that the structure the previous test found on the data can be explained solely by red noise at these timescales. Finally, all these results should consider the look-elsewhere effect related to PKS\,2155$-$305 not being chosen at random, but rather as a known quasi-periodic candidate at other timescales in other data samples; this effect, though difficult to quantify, likely overestimates the significance values, which strengthens even further the conclusion that there is no significant periodicity present in our data. We present a summary of all the PDM results in Table \ref{tab:pdmperiods}.

\begin{table}
\caption{Candidate periods derived with the PDM method for each segment. In column 1 we list the segment ID, in column 2 the candidate period, in column 3 the $\theta$ dispersion statistic derived with the PDM method, in column 4 the $p$-value derived with a surrogate test (magnitude shuffling), and in column 5 the $p$-value derived with a surrogate test considering red noise.}

\label{tab:pdmperiods}
\centering
\begin{tabular}{c c c c c}
\hline\hline
Segment  & Candidate period  & $\theta_\text{PDM}$ & $p$-value & $p$-value\\ 
 & [days] & & [mag-shuffle] & [red noise] \\
\hline
 5   & 13.6 & 0.016 & $<$1E-4 & 0.9997  \\

 8   & 29.7 & 0.015 & $<$1E-4 & 0.7661 \\

 9   & 19.6 & 0.008 & $<$1E-4 & 0.9430 \\
\hline
\end{tabular}
\end{table}

\section{Summary and conclusions}

We used 4-band optical photometric observations of the blazar PKS\,2155$-$304 spanning more than ten years to investigate its variability properties, both in flux and spectral index, across a range of timescales—from years down to a few hours. The data were analysed for signs of periodicity and for possible correlations with the source’s emission at X-ray frequencies. Our main results are summarised below.

   \begin{enumerate}
      \item This study confirms the significant optical variability of the object in time scales from days to years. Nevertheless, our data indicate that the source was in a relatively low-activity state during our monitoring, compared to the higher activity levels reported by \cite{sandrinelli14} and in ASAS-SN observations \citep{Kochanek17}, as can be clearly seen in the light curves (Figure \ref{fig:lc_comparison_literature}). It is observed that, at the longest scales, results of the variability tests were generally positive. This 
      is not the case for microvariability, since in very few cases significant flux variations were detected above the adopted confidence level. However, microvariabilities were indeed found for times shorter than 4 hours. We confirm that the $F$-test tends to yield positive results more frequently than the $C$-test, as previously noted in \citet{zibecchi2017}.
      
      \item  There was a noticeable change in the flux of the object in all 4 optical bands. For example, between July 27, 1997 and July 11, 2023 (9480 days) there was a drop measured in the $V$ band of 1.43 magnitudes. In the $R$ band a brightness drop of 0.99 magnitudes could be measured in 2509 days. While a roughly similar behaviour between photometric bands is evident, we were able to trace changes in the spectral index. %\sac{\sout{At the same time, it is observed that on the nights where data were available in different photometric bands, the behaviour of the respective magnitudes was similar, confirming the strong correlation between these bands}  }.
      \item Cross-correlation results using the ZDCF between the optical band and X-Rays for time lags less than 100 days showed positive correlations, albeit small and with large associated errors. When higher order $\tau$ results were considered, no particular pattern was noticed, so we considered the case where these 2 bands were in fact uncorrelated, which would rule out physical processes such as jet precession and imply that the AGN component responsible for the emission of both bands would not be the same with the X-ray component likely originating from the accretion disk rather than the jet.
      \item It is relevant to note that during the period between July and August 2023, the object experienced a significant increase in its average flux, both in the $R$ and $V$ bands, exceeding 0.5 magnitudes in less than 1 month. This phenomenon is consistent with the sudden increase predicted, supporting the occurrence of a flare during that epoch.
      \item The calculated spectral index was always negative, taking values consistent with a spectrum characterized by non-thermal radiation. In addition, it is evident that the spectrum of the object became flatter. The observed spectral flattening likely reflects changes in the particle acceleration or the relative contributions of emission components within the jet. Despite some evidence suggesting a mild bluer-when-brighter effect, the data scatter remains too large and the tests performed were not significant enough to support it conclusively.
      
      \item We chose to look for quasi periodicity of periods $<$\,30 days within segments of 100 days of consecutive data, to avoid yearly gaps. The PDM method yields periodicity when compared against 1E4 simulated light curves, which becomes statistically non-significant ($p>0.74$, for all cases) when these light curves are simulated with red noise. We thus claim that the structure apparent in the data can be explained solely due to red noise. Although periodic behavior has been reported in blazars and could be physically explained by mechanisms such as geometrical modulation due to jet precession \citep{britzen} or turbulence induced instabilities \citep{abdollahi}, our analysis suggests that, for PKS\,2155$-$304, such fluctuations are more likely the manifestation of a stochastic process. 
   \end{enumerate}

\section*{acknowledgements}
We thank the anonymous referee and editor for the valuable feedback that helped us improve this study.\\
E. J. M. would like to thank Dr. R. I. P\'aez, Dr. N. Masetti and Dr. R. Campana for their useful feedback on this work.\\
S. C. would like to thank R. Artola for carrying out the observations at the Bosque Alegre Observatory.\\
The present work was partially supported by grant 11/G178 from the Universidad Nacional de La Plata, Argentina.

\section*{data availability}

Based on data acquired at Complejo Astronómico El Leoncito, operated under agreement between the Consejo Nacional de Investigaciones Científicas y Técnicas de la República Argentina and the National Universities of La Plata, Córdoba and San Juan. Proposal Codes: JS-2023B-12 and JS-2024B-13. Information about the database can be found at https://casleo.conicet.gov.ar.\\ 
This work is partially based on observations obtained with the 1.54-m telescope at Estación Astrofísica de Bosque Alegre dependent on the National University of Córdoba, Argentina.\\
The data underlying this article will be shared on reasonable request to the corresponding author.\\

%%%%%%%%%%%%%%%%%%%%%%%%%%%%%%%%%%%%%%%%%%%%%%%%%%
% Bibliografía
%%%%%%%%%%%%%%%%%%%%%%%%%%%%%%%%%%%%%%%%%%%%%%%%%%
\bibliographystyle{mnras}
\bibliography{bibliografia.bib}

%\clearpage

%\section*{Appendix A: Supplementary figures and tables}
%\setcounter{figure}{0}
%\setcounter{table}{0}
%\renewcommand{\thefigure}{A.\arabic{figure}}
%\renewcommand{\thetable}{A.\arabic{table}}

%\begin{figure}
%    \centering
%    \includegraphics[width=0.4\textwidth]{lc_segment8_pdm_phase_fold.png}
%    \caption{Phase folded diagram for segment 8, implementing the period of $\sim 13.6$ days derived with the PDM method.}
%    \label{fig:pdm_phasefold_s8}
%\end{figure}

%\begin{figure}
%    \centering
%    \includegraphics[width=0.4\textwidth]{lc_segment9_pdm_phase_fold.png}
%    \caption{Phase folded diagram for segment 9, implementing the period of $\sim 29.1$ days derived with the PDM method.}
%    \label{fig:pdm_phasefold_s9}
%\end{figure}

\bsp	% typesetting comment
\label{lastpage}

\end{document}